\documentclass[aps,prc,preprint,groupedaddress,superscriptaddress]{revtex4-1}
\usepackage[utf8]{inputenc}
\usepackage{graphicx}
\usepackage{amsmath, amssymb, bm, mathrsfs, gensymb}

\begin{document}

\title{Single Charge-Exchange Reactions and the Neutron Density at the Surface 
of the Nucleus}

\author{Bui Minh Loc} 
\affiliation{School of Physics and Astronomy, Tel Aviv University, Tel Aviv 
69978, Israel.}
\affiliation{Department of Physics, Ho Chi Minh City University of 
Pedagogy, 280 An Duong Vuong Street, District 5, Ho Chi Minh City, Vietnam.}
\author{Naftali Auerbach} 
\affiliation{School of Physics and Astronomy, Tel Aviv University, Tel Aviv 
69978, Israel.}
\author{Dao T. Khoa} 
\affiliation{Institute for Nuclear Science and Technology, VINATOM 179 Hoang 
Quoc Viet Rd., Hanoi, Vietnam.}

\date{\today}

\begin{abstract}
In this work we study the charge-exchange reaction to Isobaric Analog State 
using two types of transition densities. We show that for projectiles that do 
not probe the interior of the nucleus but mostly the surface of this nucleus, 
distinct differences in the cross-section arise when the two types of transition 
densities are employed. We demonstrate this by considering the ($^3$He,$t$) 
reaction.
\end{abstract}



\maketitle

\section{Introduction \label{Intro}}
Single Charge-Exchange (SCX) reactions were and are now an excellent source of 
information about isovector properties of nuclei. In particular successful is 
the SCX to the isobaric analog state (IAS). In this process one is able to probe 
the distribution of the isovector nuclear density. The IAS is defined as:
\begin{equation}\label{ias1}
 | A \rangle = \frac{1}{\sqrt{2T}}T_- | \pi \rangle,
\end{equation}
where $| A \rangle$ denotes the IAS, $| \pi \rangle$ the parent state with 
isospin $T$, and $T_-$ is the isospin lower operator. 
The transition density for this model state is given by $\rho_n(r) - \rho_p(r)$ 
the difference between the neutron and proton densities. The Coulomb interaction 
of the protons does affect the distribution of the $Z$ protons in the nucleus 
and the density distribution of the $Z$ neutrons is different from the 
distribution of the $Z$ protons. As discussed in the past \cite{Auer83}, 
because of the Coulomb repulsion, the $Z$ protons have a larger radius compared 
to the corresponding $Z$ neutrons. The $Z$ neutrons and the $Z$ protons 
are denoted as the core (we assume that we deal with nuclei that $N>Z$). We 
make the following decomposition:
\begin{equation}
 \rho_n(r) - \rho_p(r) = \rho_{n(exc)}(r) + \delta\rho(r),
\end{equation} 
where $\delta\rho(r)$ denotes the density of the $Z$ neutrons of the core minus 
the density of the protons
\begin{equation}
 \delta\rho(r) = \sum_{i=1}^{Z} |\varphi_i^n(r)|^2 - \sum_{i=1}^{Z},
|\varphi_i^p(r)|^2
\end{equation}
and $\rho_{n(exc)}$ the density of $N-Z$ excess neutrons
\begin{equation}
 \rho_{n(exc)}(r) = \sum_{i=Z+1}^{N} |\varphi_i^n(r)|^2,
\end{equation}
with $\varphi_i^{n(p)}(r)$ being the neutron (proton) single-particle wave 
function. The volume integral of $\delta\rho(r)$ must be zero and therefore 
this term must have at least one node. The inside part is positive while the 
surface part is negative because there is an excess of protons outside, since 
the protons are expelled by the Coulomb interaction. The $\delta\rho(r)$ term 
was studied in the past and it was shown that the shape of this distribution can 
be approximated by the relation \cite{Auer83}:
\begin{equation}
 \delta \rho (r) \sim \left(3 \rho(r) + r\frac{d\rho(r)}{dr}\right),
\end{equation} 
where $\rho(r)$ is the total nuclear density.

\begin{figure}[!t]
\begin{center}
\includegraphics[scale=0.6]{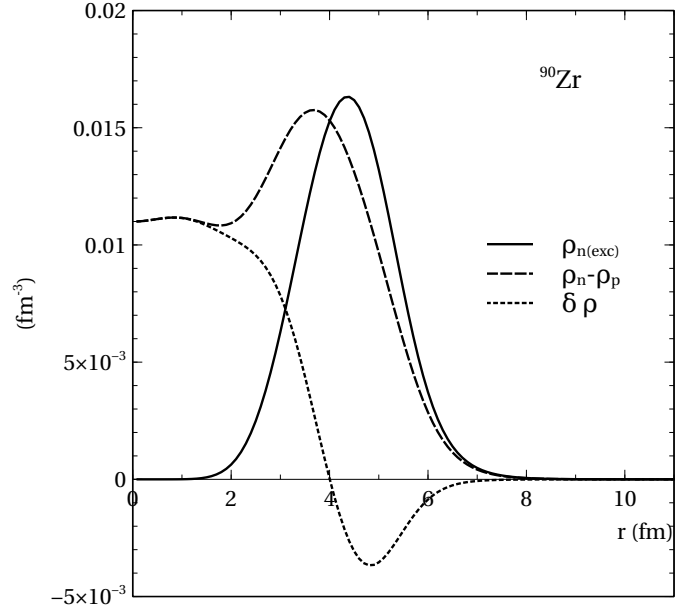}
\caption{$\rho_n(r) - \rho_p(r)$, $\rho_{n(exc)}(r)$ , and $\delta \rho(r)$ of 
the $^{90}$Zr nucleus obtained from the HF-BCS calculation using the BSk17 
version of Skyrme interaction. \label{90ZrDens}}
\end{center}
\end{figure}
\begin{figure}[!t]
\begin{center}
\includegraphics[scale=0.6]{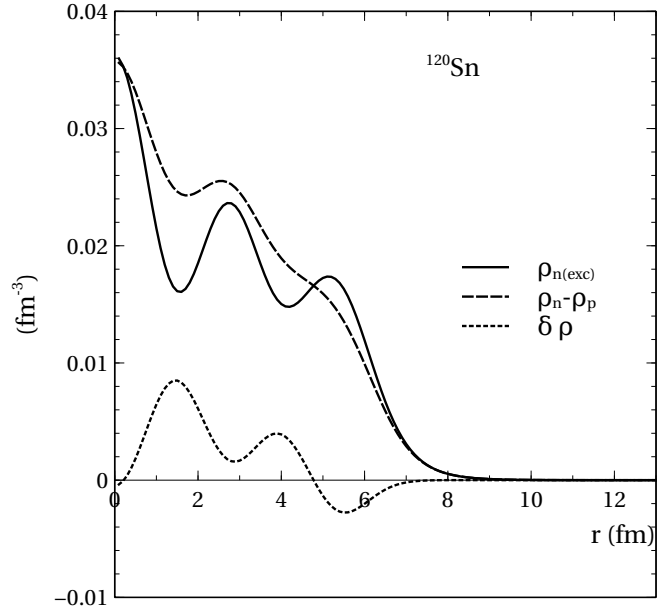}
\caption{The same as in Fig. \ref{90ZrDens}, but for $^{120}$Sn nucleus. 
\label{120SnDens}}
\end{center}
\end{figure}
\begin{figure}[!t]
\begin{center}
\includegraphics[scale=0.6]{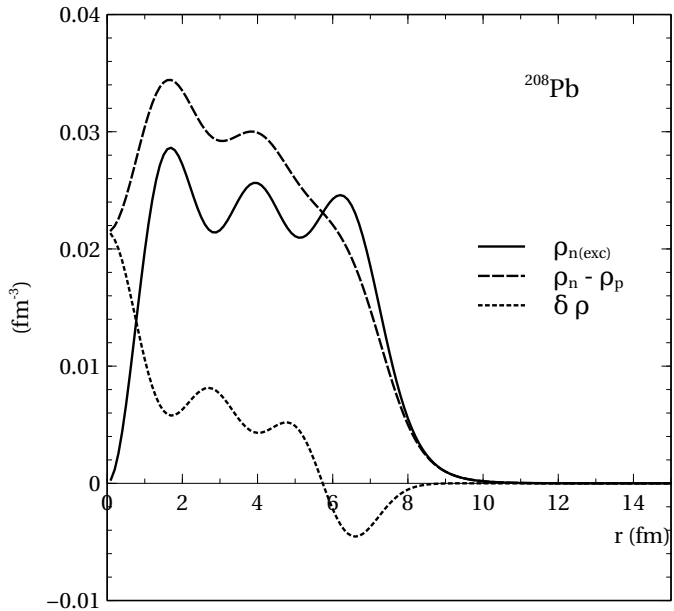}
\caption{The same as in Fig. \ref{90ZrDens}, but for $^{208}$Pb nucleus. 
\label{208PbDens}}
\end{center}
\end{figure}

When a projectile probes the interior of the nucleus it will experience the 
interior transition density as well as the exterior one. These are of opposite 
signs in $\delta \rho(r)$ and therefore there \textit{should not be much 
difference between the use of $\rho_n(r) -\rho_p(r)$ or $\rho_{n(exc)}(r)$}. 
This changes when the projectile reaches the surface but does not penetrate the 
interior. In this case the projectile will experience the excess neutron density 
and the external part of $\delta \rho(r)$. This really means the 
projectile experience somewhat less neutrons at the surface compared to the case 
when only excess neutrons are present. Thus the two transition densities will 
\textit{give different results for the SCX cross-sections to the IAS}. The 
transition will be, therefore, larger when the excess neutron density is used. 
In the past this picture was shown to be valid for pion SCX reactions were 
used. We are still left with the question, which of the two transition densities 
should be used in the SCX to the IAS. This question was answered in the past in 
several references \cite{Auer81, Auer82}. 

If one uses the definition of the IAS, Eq. (\ref{ias1}), then $\rho_n(r) - 
\rho_p(r)$ is the correct one. But this is not the physical analog state. It was 
shown in \cite{Auer81, Auer82} that due to the Coulomb interaction the physical 
state is such that the $\rho_{n(exc)}$ is the proper transition density. When 
the $T_-$ operator acts on all neutrons it also affects the core neutrons 
because the corresponding proton orbits are slightly different form the neutron 
orbits and thus the Pauli principle allows partially to change the neutron wave 
functions when the $T_-$ operator acts. However the physical IAS does not have 
the core affected. $^{41}$Ca ground state and its IAS, that is the ground state 
of $^{41}$Sc have the same cores, only the last neutron, with a neutron wave 
function is transformed into a proton in the same orbit but with a proton wave 
function. A correct description of this is to use the \textit{analog spin 
scheme} \cite{Mekj68, Auer81} in which the $W_-$ operator changes a neutron with 
a neutron wave function into a proton in the same orbit but with the proton wave 
function.

In the recent years, new SCX to the IAS experiments were performed using light 
ions, in particular the ($^3$He,$t$) reaction. Also the theoretical 
analysis of these reactions have been presented \cite{Loc14}.

\section{Method of Calculation\label{Method}}
The method of calculation to obtained the differential cross-section in this 
work is as same as Ref.~\cite{Loc14}. The SCX to the IAS 
is described within the distorted wave Born approximation (DWBA).
The phenomenological optical parameters of the $^3$He scattering from $^{58}$Ni 
and $^{90}$Zr are taken from Ref.~\cite{Kami03}. For $^{208}$Pb target, the 
parameters are taken from the optical model fit \cite{Zege07} of the elastic 
$^3$He scattering data at 450 MeV \cite{Yama95}. The SCX form factor is given 
by the double-folding model (DFM) in the following form     
\begin{eqnarray}\label{Fcx}
 F_{\rm cx}(R) = \sqrt{\frac{2}{T}}\int\!\!\!\int 
 [\rho_n^a(\bm{r}_a) - \rho_p^a(\bm{r}_a)] t_{01}(E,s) [\rho_n^A(\bm{r}_A) - 
\rho_p^A(\bm{r}_A)] d\bm{r}_a d\bm{r}_A,
\end{eqnarray}
where $s$ is the relative coordinate between a nucleon in projectile and a 
nucleon in the target. The nucleon-nucleon effective interaction $t_{01}(E,s)$ 
in Eq. (\ref{Fcx}) is the Franey-Love $t$-matrix \cite{Love81, Fran85}. The 
neutron and proton densities of $^3$He are given by the microscopic three-body 
calculation \cite{Niel01} using the Argonne nucleon-nucleon potential. The 
calculations of the densities and radii of target nuclei are performed using the 
Hartree-Fock (HF) \cite{Colo13} or in cases of open shell nuclei using HF-BCS 
approximations \cite{Colo17} with Skyrme type interactions. The BSk17 
parametrization 
\cite{Gori09} was employed. The details of folding model calculation for SCX 
reaction to the IAS was given in Ref.~\cite{Khoa14}. The DWBA calculations were 
done with the relativistic kinematics, using the code ECIS06 written by Raynal 
\cite{ECIS06}.

\begin{table}[b]
\caption{Properties of nuclear densities calculated using the Skyrme HF-BCS 
calculation. \label{tab1}}
\centering
\begin{ruledtabular}
\begin{tabular}{c|c|c|c|c}
& $^{58}$Ni & $^{90}$Zr & $^{120}$Sn & $^{208}$Pb \\
\hline
$(N-Z)/A$ & 0.034     & 0.111     & 0.167      & 0.212 \\
$r_n$ & 3.691     & 4.267     & 4.706      & 5.594 \\
$r_p$ & 3.694     & 4.202     & 4.573      & 5.441 \\
$r_n - r_p$ & -0.003    & 0.065     & 0.133      & 0.153 \\
$(r_n - r_p)_{core}$ & -0.047 & -0.103 & -0.070 & -0.130 \\
$r_{n(exc)}$ & 4.249  & 4.882  &  5.174 & 6.086 \\
$S_{\delta\rho(sur)}$ & -0.531 & -1.691 & -1.453 & -3.683 \\
$S_{\delta\rho(sur)}/(N - Z)$ & -0.258 & -0.169 & -0.072 & -0.083 \\
\end{tabular}
\end{ruledtabular}
\end{table}

\begin{figure}[!ht]
\begin{center}
\includegraphics[scale=0.7]{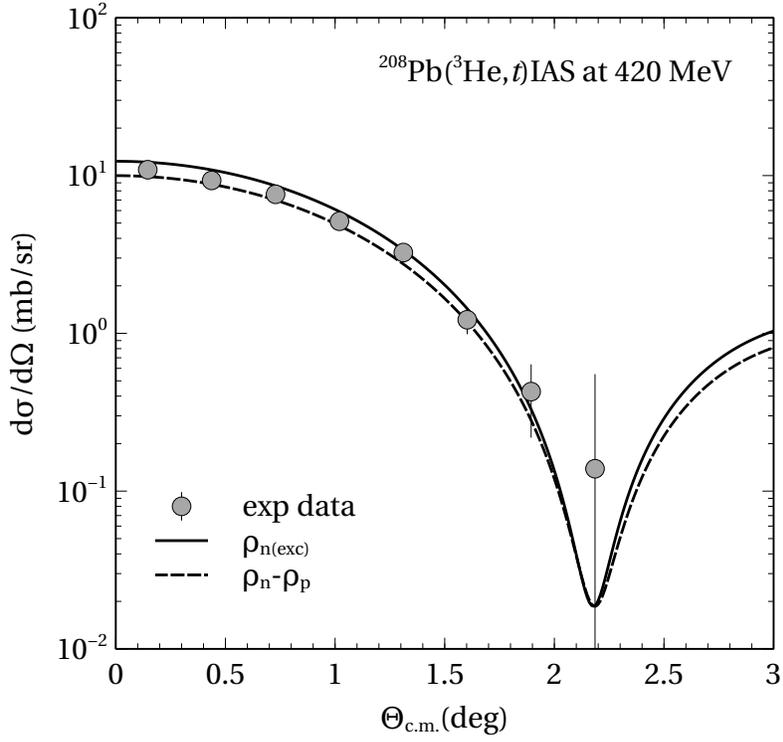}
\caption{Differential cross section of the $^{208}$Pb($^3$He,$t$) reaction 
to the IAS at $E_{\rm lab}=420$ MeV, given by the DWBA calculation using the 
SCX form factors obtained with the $\rho_n(r) - \rho_p(r)$ (dashed curve) and 
the $\rho_{n(exc)}(r)$ (solid curve). The experimental data were taken from 
Ref.~\cite{Zege07}. \label{208Pb3Het420}}
\end{center}
\end{figure}

\section{Results and Discussion \label{Results}}
In describing the results we do it in two steps. First we present the 
results of the structure calculations as these are the input in the reaction 
computations. In the second step we show the cross section for the two 
reactions ($^3$He,$t$) and ($p,n$).

\begin{figure}[!ht]
\begin{center}
\includegraphics[scale=0.7]{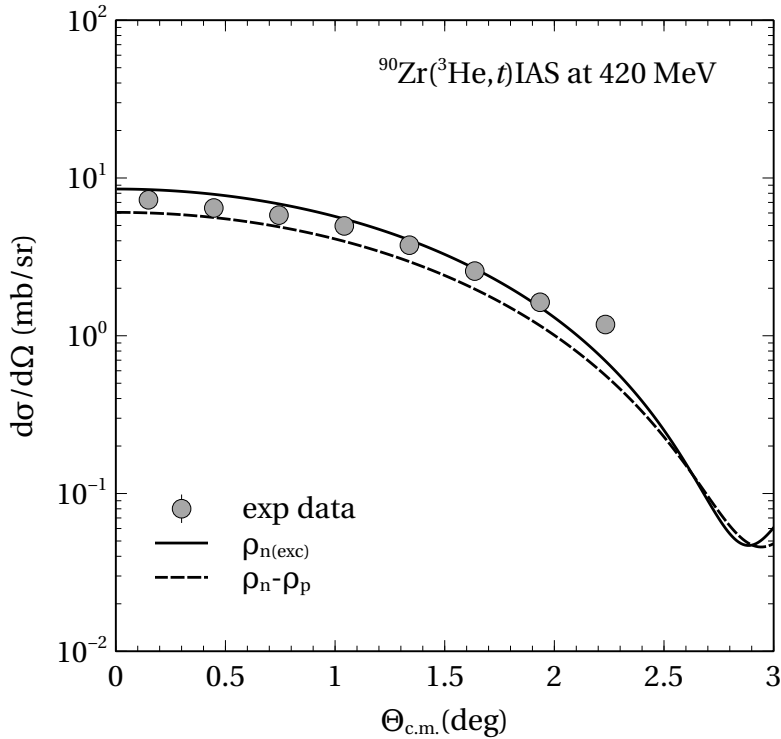}
\caption{The same as in Fig. \ref{208Pb3Het420}, but for $^{90}$Zr target. 
\label{90Zr3Het420}}
\end{center}
\end{figure}

The calculated densities $\rho_n(r) - \rho_p(r)$, $\rho_{n(exc)}(r)$ , and 
$\delta \rho(r)$ for several nuclei are shown in Fig. \ref{90ZrDens}, 
\ref{120SnDens}, and \ref{208PbDens}. The curve of $\rho_n(r) - \rho_p(r)$ is 
the sum of $\rho_{n(exc)}(r)$ and $\delta\rho(r)$. One sees that $\delta\rho(r)$ 
has a node and the inner region is positive meaning that there are more neutrons 
than protons in the $N=Z$ core but in the outer region the density is negative, 
thus there is a surplus of protons, due to the Coulomb repulsion. It is clear 
that when one uses the $\rho_n(r) - \rho_p(r)$ transition density one has less 
neutrons at the surface then in the case when the transition density is 
$\rho_{n(exc)}(r)$. The effect will be the largest when there are fewer excess 
neutrons in the nucleus as is the case of $^{58}$Ni. This will affect the SCX 
reactions when the projectiles are absorbed more strongly and do not reach the 
interior, in particular ion projectiles such as $^3$He. When a projectile 
traverses the entire (or most) of the nucleus this effect of Coulomb 
polarization density $\delta\rho(r)$ will be small. One should 
expect therefore that in ($p,n$) reactions the effect of $\delta\rho(r)$ will be 
less pronounced to that of ($^3$He,$t$) reaction (of course this also depends 
on the energies of the projectiles, as for different energies the  absorption 
might be different). In the past it was pointed out \cite{Auer98, Auer89} and 
also confirmed experimentally \cite{Prou00} that projectiles that are strongly 
absorbed will excite states that have radial transition densities consisting of 
a volume and surface parts of opposite sign, as for example the giant monopole 
or spin monopole \cite{Auer98, Auer89, Prou00}. 

In Table \ref{tab1} some of the properties of the densities and radii of nuclei 
in the study are summarized. The meaning of various quantities appearing in the 
table is obvious except $S_{\delta\rho(sur)}$ which denotes the integral of 
$\delta\rho(r)$ from its last node $R_s$ of the density to infinity,
\begin{equation}
 S_{\delta\rho(sur)} = 4\pi\int_{R_s}^\infty [(\rho_n(r) - \rho_p(r)) - 
\rho_{n(exc)}(r)] r^2 dr.
\end{equation} 
This is an illustrative, approximate way to quantify the amount of protons that 
are at the surface due to the Coulomb polarization of the $Z$ protons in the 
core. As already mentioned the effect of $\delta \rho (r)$ is largest when the 
number of excess neutrons is small. Note that in $^{58}$Ni the difference $r_n 
- r_p$ is actually negative. This is in agreement with the prediction in Ref. 
\cite{Auer10} and the difference $(r_n - r_p)_{core}$ is in reasonable 
agreement with the formula derived in the Ref.~\cite{Auer10}
\begin{equation}
 (r_n - r_p)_{core} = -1.6 \times 10^{-3} Z.
\end{equation}

So what effect do the above density distribution have on the SCX cross-section? 
We now discuss the results of the DWBA calculations. The structure ingredients 
discussed above are tested in our analysis of the ($^3$He,$t$)IAS.
In Fig. \ref{208Pb3Het420} the ($^3$He,$t$) differential cross-sections 
at 420 MeV are shown for $^{208}$Pb using the transition densities $\rho_n(r) - 
\rho_p(r)$ and $\rho_{n(exc)}$. We see that the cross-section calculated with 
$\rho_{n(exc)}(r)$ is slightly higher than with $\rho_n(r) - \rho_p(r)$ but the 
difference is not large, consistent with the fact that the number of protons 
pushed out by the Coulomb force is large (about 4) but compared to the 44 excess 
neutrons the effect is small. 

\begin{figure}[!t]
\begin{center}
\includegraphics[scale=0.7]{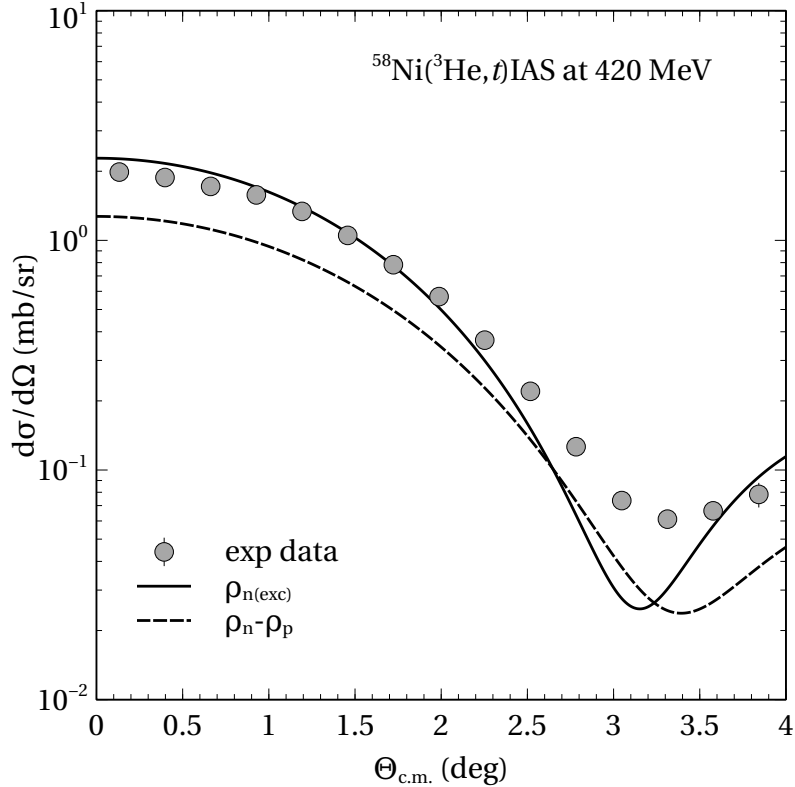}
\caption{The same as in Fig. \ref{208Pb3Het420}, but for $^{58}$Ni target. 
\label{58Ni3Het420}}
\end{center}
\end{figure}

In Fig. \ref{90Zr3Het420} the same results are plotted for $^{90}$Zr but in 
this case due to the smaller number of excess neutrons the effect is larger. The 
cross-section with $\rho_{n(exc)}(r)$ is larger and closer to the experimental 
results \cite{Zege07}. The results for $^{58}$Ni are shown in Figure 
\ref{58Ni3Het420}. Here the number of excess neutrons is 2, and the effect of 
$\delta \rho_{sur}$ compared to $\rho_{n(exc)}$ is sizable (see Table 
\ref{tab1}). The cross-section in the forward direction is increased by more 
than a factor of 2, agreeing with the experimental data.

\begin{figure}[!ht]
\begin{center}
\includegraphics[scale=0.7]{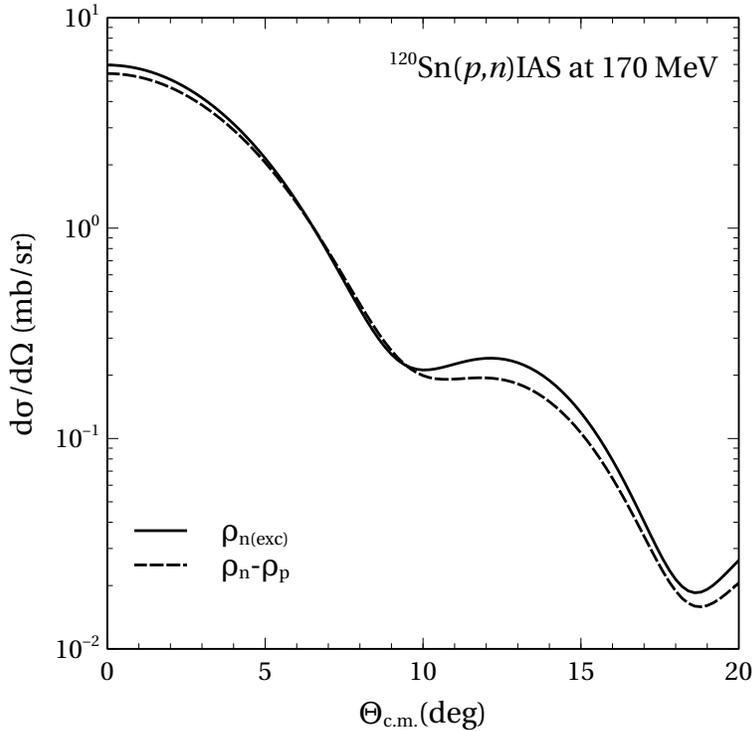}
\caption{Differential cross sections of the $^{120}$Sn($p,n$)IAS reaction 
at $E_{\rm lab}=170$ MeV, given by the DWBA calculation using the SCX 
form factors obtained with $\rho_n(r) - \rho_p(r)$ (dashed curve) and 
the $\rho_{n(exc)}(r)$ (solid curve). \label{120Snpn170}}
\end{center}
\end{figure}

It is interesting to contrast the ($^3$He,$t$)IAS reaction with the 
($p,n$)IAS reaction. As mentioned above the latter one (depending on the 
energy) may probe the interior of the nucleus and would be less sensitive to 
the polarization of the core. The results of the calculations using the two 
transition densities should be close. In Fig. \ref{120Snpn170}, we show the 
prediction for the $^{120}$Sn($p,n$)IAS reaction at 170 MeV for the two 
transition densities. Here the difference between the two curves is very small.

\section{Conclusions}
We discussed in the present work the impact of two forms of 
transition densities used in the charge-exchange reactions to the IAS. We found 
that when the projectile used in the reaction does not probe the interior of the 
nucleus but mostly the surface, visible difference in the cross-sections arise 
when the two densities are employed. The ($^3$He,$t$)IAS reaction at medium 
energies is of this type and in nuclei with a low number of excess neutrons this 
effect is enhanced. Single charge-exchange, and double charge-exchange reactions 
with complex projectiles may provide useful tools to study the neutron-proton 
content at the surface of the nucleus.

\begin{acknowledgments}
The authors thank to Gianluca Col\`{o} for providing us with the HF-BCS code. 
This work was supported, in part, by the US-Israel Binational Science 
Foundation (grant 2014.24) and Vietnam's National Foundation for Science and 
Technology Development (NAFOSTED project No. 103.04-2014.76).
\end{acknowledgments}

\end{document}